\documentstyle[prl,aps]{revtex}

\begin{document}
\preprint{v1}
\draft

\twocolumn[\hsize\textwidth\columnwidth\hsize\csname
@twocolumnfalse\endcsname
\title{Quantum self-consistency of $AdS \times \Sigma$ brane models.}
\author{Antonino Flachi\footnote{flachi@ifae.es}, Oriol Pujol{\`a}s\footnote{pujolas@ifae.es}}
\address{IFAE, Campus UAB, 08193 Bellaterra (Barcelona), Spain}
\date{\today}
\maketitle

\begin{abstract}
    Continuing on our previous work, we consider a class of higher dimensional
    brane models with the topology of $AdS_{D_1+1} \times \Sigma$, where
    $\Sigma$ is a one-parameter compact manifold and two branes of codimension one
    are located at the orbifold fixed points. We consider a set-up where 
    such a
    solution arises from Einstein-Yang-Mills theory and evaluate the one-loop
    effective potential induced by gauge fields and by a generic bulk
    scalar field.
    We show that this type of brane models resolves the gauge hierarchy between
    the Planck and electroweak scales through redshift effects due to the warp
    factor $a=e^{-\pi kr}$. The value of $a$ is then fixed by minimizing
    the effective potential.  We find that, as in the Randall Sundrum case, 
    the gauge field contribution to the effective potential
    stabilises the hierarchy without fine-tuning as long as the laplacian
    $\Delta_\Sigma$ on $\Sigma$ has a zero eigenvalue. Scalar fields can
    stabilise the hierarchy depending on the mass and the non-minimal 
    coupling.  
    We also address the quantum self-consistency of the
    solution, showing that the classical brane solution is not spoiled by
    quantum effects.\\

\end{abstract}
\pacs{~ \hfill }
\hfill \vbox{\hbox{UAB-FT-543}}

\vskip2pc]

\newcommand{\MM}{\hat{M}}
\newcommand{\LL}{\hat{\Lambda}}
\newcommand{\k}{l}
\newcommand{\nn}{\nonumber\\}
\newcommand{\hsi}{\hat{\sigma}}
\newcommand{\hrho}{\hat{\rho}}
\newcommand{\hl}{\hat{\lambda}_\k}
\newcommand{\si}{\sigma}
\newcommand{\nul}{\nu_l}
\newcommand{\beq}{\begin{equation}}
\newcommand{\eeq}{\end{equation}}
\newcommand{\bed}{\begin{displaymath}}
\newcommand{\eed}{\end{displaymath}}
\def\bea{\begin{eqnarray}}
\def\eea{\end{eqnarray}}

\section{Introduction} 
\label{introduction}
An original way to address the gauge hierarchy problem has been suggested by
Randall and Sundrum, by considering five dimensional anti de Sitter spacetime
compactified on $S^1/Z_2$ and two $3-$branes located at the or\-bi\-fold fixed
points \cite{rs}.  This set-up results in a non factorisable geometry, which has
the virtue of resolving the large ratio between the Planck and electroweak
scales as a geometrical effect: in the four dimensional effective theory
the masses on the negative tension brane are redshifted by a factor $a=e^{-kr\pi}$ ($1/k$ is the
curvature of anti de Sitter spacetime and $r$ is the radius of the orbifold).
Hence, it is possible to generate a TeV mass scale from a Planck sized mass
by taking $kr \sim 12$.

Obviously, to make this scenario consistent, the size of the orbifold has to be
determined dynamically and not fixed by hand.  In this sense, the RS model does
not completely solve the hierarchy problem, unless a stabilisation mechanism for the size of the
fifth dimension is included. (It is worth noting that this is not a peculiarity
of the RS model, rather a well known feature of higher dimensional 
theories with extra dimensions.)

Within the RS proposal, a way of achieving such a stabilisation was initially
suggested by Goldberger and Wise, introducing an appropriate classical
interaction between the branes and a bulk scalar field \cite{gw}.  In this way,
it is possible to stabilise the extra fifth dimension without fine tuning,
however, the lack of a fundamental origin for such interaction renders such a
mechanism artificial.

An alternative to the GW mechanism can, in principle, be the Casimir energy
generated by quantum fields. Already in the old Kaluza-Klein theories, Candelas
and Weinberg looked at this possibility \cite{cw}, and, inspired by their work,
several authors investigated the role of quantum effects in the new brane models
\cite{gpt1,flachi1,flachi3,flachi2,golrot,gpt2}.
In particular, as a result of such studies, it has recently been realized \cite{garpom}
that the quantum effective potential due to bulk gauge fields can stabilise the
radion and generate the hierarchy of scales without fine tuning. This provides a
viable alternative to the GW mechanism in the RS model.
Further aspects of quantum effects have been investigated in
\cite{wade,nojiri1,ogushi,mukohyama,wade&misao1,wade&misao2,hofmann,brevik,ponton,saharian}.

The brane world idea has opened up a range of interesting possibilities in
addressing many long standing problems of particle physics and cosmology.  As
one of the prototypes, the RS scenario can be viewed as a model belonging to a
larger class and especially in connection with a possible embedding of such
scenarios within string theory, it is worth exploring at more depth extensions
of the RS model by considering spacetimes with higher dimensionalities and
curved internal spaces (Explicit six dimensional examples can be found in
\cite{gherghetta,gregory,ghrsh,oda,kanti,kogan}.). Needless to say, such
extensions provide models with a richer structure than the RS one (See
\cite{rizzo,shap}).

Many generalisations of the Randall-Sundrum model fall in the quite general
class of higher dimensional warped solutions studied in \cite{daemi}, where a
$D-$dimensional system of gravity plus Yang-Mills is considered. The base
spacetime is described by the following line element:
\begin{equation}
    \label{metric}
    d\hat{s}^2= e^{2\sigma(y)}\eta_{\mu\nu}dx^\mu dx^\nu
    + e^{2\rho(y)} g^{\Sigma}_{ij}dX^idX^j + dy^2 ,
\end{equation}
where the coordinates $x^\mu$ parametrise $D_1$ dimensional Minkowski space
${\cal M}$, the coordinates $X^i$ cover a $D_2-$dimensional compact internal
manifold $\Sigma$ of radius $R$
and the coordinate $y \in [-\pi r, \pi r]$ parametrises the orbifold. We define
$D=D_1+D_2+1$ and take $D_1=4$. Hatted quantities refer to higher dimensional
ones.

Depending on the geometry of the internal space, Einstein equations lead to
different types of warp factors $\sigma(y)$ and $\rho(y)$: when the internal
space is a Ricci flat manifold and the Yang-Mills flux is switched off, the
general result for the warp factors is given by a combination of exponentials.
Simpler solutions with \beq \sigma(y)=\rho(y)=-k|y|,
\label{type1}
\eeq are found when the bulk cosmological constant is taken to be negative.
Additionally, the condition of Ricci flatness of $\Sigma$ can be relaxed at the
price of introducing some extra bulk matter, like, for instance, a scalar field
with hedgehog configuration \cite{ghrsh,oda}. In the above cited papers, the
set-up allows for the presence of one brane only; two brane models can be
constructed by gluing two slices of the previous spacetime and imposing the
$Z_2-$identification.

Solutions of the type $AdS_{D_1+1}\times \Sigma$ are, instead, found when the
internal space is non Ricci flat without adding any extra bulk matter. In such
case the warp factor along $\Sigma$ is constant and we can take\beq \rho(y)=0,
\sigma(y)=-k|y| .
\label{type2}
\eeq 
We note that in such case the requirement of a negative higher dimensional
cosmological constant can be relaxed.

In a previous paper \cite{flachi4}, we have considered scenarios of the type
(\ref{metric}), (\ref{type1}) and evaluated the effective potential arising from
massive, non-minimally coupled scalar fields. We showed that one-loop effects
generate a suitable effective potential which can stabilise the hierarchy
without fine tuning, provided that the internal space is flat. Aside, we have
also seen that considering a warped internal manifold was providing a novel way
to solve the hierarchy, which was relying on a mixture of large volume and
redshift effects.

Here, we wish to extend the previous results to the second type of spacetimes
(\ref{metric}), (\ref{type2})\footnote{In passing, it is worth noting that this
  type of metrics arise in string theory. Here we won't be concerned with any
  string theory application, and simply refer the reader to references
  \cite{gauntlett,cvetic}, where this type of solutions are found in the more
  fundamental context of supergravity or M-theory.}.

The plan of the paper is the following.  In the next section we briefly discuss
the model and comment on the relevant mass scales of our set-up. Section~\ref{2}
is devoted to compute the quantum effective potential arising from a quantised
bulk scalar and gauge field. The possibility of stabilising the hierarchy by
using quantum effects is discussed in section~\ref{4}, where the problem of the
quantum self-consistency of the solution is also addressed. In the last section
we report our conclusions.  Some results concerning the uniform asymptotic
expansion of the Bessel functions (needed in the computation of the effective
potential) are collected in appendix.

\section{Background solution and scales} 
\label{solution}

In the present section we describe the background solution when branes are
included and discuss the relevant scales of the problem.

The spacetime we consider corresponds to the line element (\ref{metric}) where
the warpings satisfy the condition (\ref{type2}). In other words, the bulk
spacetime we consider is the direct product of five dimensional anti de Sitter
space and a compact one-parameter manifold $\Sigma$.

Randjbar-Daemi and Shaposhnikov have considered this type of solutions and
showed that they arise from a system of gravity plus Yang-Mills fields
\cite{ghrsh,daemi}, with bulk action given by \beq S_{BG}=\int d^Dx
\sqrt{\hat{g}} \left\{ \hat{M}^{D-2} \hat{{\cal R}} - \hat{\Lambda} -{1\over
      4\hat{g}_*^2} \hat{{\cal F}}_{IJ}\hat{{\cal F}}^{IJ} \right\}~.
\label{bulk}
\eeq

The equations of motion can be obtained in the standard way, and once the ansatz
for the metric tensor (\ref{metric}),  
(\ref{type2}) is used, the following independent equations are obtained:
\beq
k^2 = -{\MM^{2-D}\LL\over D_1(D-2)} + {\MM^{2-D}\over D_1(D-2)}{\hat{F}^2\over
  4\hat{g}_*^2R^4}~,  
\label{eqmunu}
\eeq
\beq
{\Omega\over R^2}={\MM^{2-D}\over D-2} \left\{\LL+{2D-D_2-4\over
      D_2}{\hat{F}^2\over 4\hat{g}_*^2R^4}\right\}~, 
\label{eqij}
\eeq
where we have expressed the curvature ${\cal R}_{\Sigma}$ of the internal
manifold in terms of its radius $R$ ($\Omega$ is a constant): 
$$
{\cal R}_{\Sigma}= {D_2 \Omega \over R^2}~,
$$
and 
$$
{\hat{F}^2 \over R^4} = \hat{g}^{IM}\hat{g}^{JN}\hat{{\cal F}}_{IJ}\hat{{\cal
    F}}_{MN}~. 
$$
The previous equations (\ref{eqmunu}), (\ref{eqij}) allow us to determine the
radius of the internal manifold and the Yang-Mills flux in terms of $\LL$, $\MM$
and $k$: 
\bea
\label{erre}
R^2 &=& \Omega D_2 {\cal P}^{2}~, \\ 
{\hat{F}^2\over 4\hat{g}_*^2} &=&
D_2^2 \Omega^2 {\cal P}^{4}
( \LL + D_1(D-2)k^2 \MM^{D-2})~,
\label{effe}
\eea
where, for notational convenience, we have defined
$$
{\cal P}^{-2} = 2\MM^{2-D}\LL +D_1(2D-D_2-4)(D-2)k^2.
$$
One immediately notices that the radius of $\Sigma$ is `stabilised' at
classical level at the price of tuning the Yang-Mills flux according to
(\ref{effe}).  In this sense we point out some analogy with the recent work of
Carroll and Guica \cite{carroll}, who considered the direct product of Minkowski
space and a 2-sphere. In their case the radius of the 2-sphere is stabilised by
the flux and a relaxing the tuning of such flux would induce a de Sitter or
anti-de Sitter geometry rather than Minkowski. The same is also true in our case
with the additional modification of the warp factor.

Since we are considering two branes embedded in such a spacetime, we have to add
to the action appropriate brane tension terms. It is easy to see that there are
no solutions of the type considered here, if the tension term is isotropic. 
The requirement of conservation of the higher
dimensional energy-momentum tensor along with the junction conditions forces us
to introduce such anisotropy\footnote{The source for such anisotropy can be due
  to different contributions to the vacuum energy or also due to a background
  three-form field \cite{chen}.} \cite{kanti,kogan,chen}.  

The brane energy-momentum tensor is then given by:
\bea
T_\mu^\nu &=&  
\delta(y) ~\mbox{diag} \left(\tau_-^{\cal M}
    \delta_\mu^\nu~,~~\tau_-^{\Sigma}\delta_i^j \right) + \nonumber \\ 
&+& \delta(y-\pi r)~\mbox{diag} \left( \tau_+^{\cal M}
    \delta_\mu^\nu~,~~\tau_+^{\Sigma}\delta_i^j \right)~. 
\eea
The spacetime we are considering can then be constructed by gluing two copies of
a slice of the bulk space and imposing  
the $Z_2-$identification. The Israel junction conditions fix the brane tensions
to be 
\beq
\tau^{\cal M}_{\pm}= {D_1-1\over D_1}\tau^{\Sigma}_{\pm} =
{D_1 -1\over D_1}\left( \mp 4D_1k\MM^{D-2} \right)~.
\eeq

We can now look at the physical scales to see whether such class of models
suggests anything about the gauge hierarchy problem.

By integrating out the extra dimensions we can write a relation between the
four- and higher-dimensional Planck scales 
\beq
m_{P}^2 = {v_\Sigma \over D_1 -2} (\MM R)^{D_2} {\MM\over k} \MM^{D_1 -2}~, 
\eeq
where
$$
v_{\Sigma}R^{D_2}= \int_\Sigma  \sqrt{g^{\Sigma}}d^{D_2}X~
$$
and the EW/Planck hierarchy can then be written, for $D_1=4$, as
\beq
h^2\equiv a^2{\MM^2\over m_P^2} \sim {a^2 \over (R\MM)^{D_2}}{k\over \MM} \sim 10^{-32}~.
\label{h}
\eeq

We see that, analogously to \cite{flachi4}, the hierarchy $h$ is expressed in
terms of $a$ and $R$, however in the present case it is not possible to use both
the redshift and large volume effects as in our previous work \cite{flachi4}.
To see this, we remind that in the case of equal warpings the crucial ingredient
was that the internal manifold was growing exponentially away from the negative
tension brane located at $y=y_-$ and this was diluting gravity as in models with
large extra dimension. On the other hand, gauge interactions, confined on the
negative tension brane, were not diluted because the size of $\Sigma$ at $y=y_-$
was of order of the fundamental cut-off.

Here the situation is different as we are considering the direct product $AdS
\times \Sigma$.  In such case, the size, $R$, of the internal manifold has to be
everywhere small, if we require that the extra $\Sigma-$dimensions remain
invisible to ordinary matter.  Since $R$ is determined at
classical level, Einstein equations leave us with a first `consistency' check on
such class of models if we were going to construct any
(pseu\-do-)\-rea\-lis\-tic scenario.

If we express the cosmological constant by factoring out two powers of the mass,
\beq
\LL \sim \lambda^2 \MM^{D-2}~,
\eeq 
relations (\ref{erre}), (\ref{effe}) can be recast in the following form:
\bea
{\hat{F}^2\over 4\hat{g}_*^2} &\sim& {\MM^{D-2} \over (k^2+\lambda^2)}~, \\
R^2 &\sim& {1\over k^2+\lambda^2}~.
\eea
Now, a natural assumption is that the bulk cosmological constant is of the same
order as the higher dimensional Planck scale,  
$\lambda \sim \MM$, and $k$ smaller than $\MM$, implying
\beq
R \sim \MM^{-1}~,
\label{R}
\eeq
meaning that the size of the internal manifold is of order of the cut-off and
thus satisfying the requirement of small $R$. The previous relation also implies
that 
\beq
(kR)^2 \sim {k^2 \over \lambda^2 + k^2} << 1~.
\eeq
This last condition will be tacitly used in the subsequent computation of the
effective potential. 

From the gauge hierarchy point of view, this class of models does not suggest
any improvement with respect to the RS model. As one can see from (\ref{h}) and
(\ref{R}), the hierarchy is resolved only through redshift effects.  Obviously,
one could relax the condition $\lambda \sim \MM$, but this in turn would
interchange the gauge hierarchy problem with the need for an `ad hoc' tuning of
the bulk cosmological constant, as we would have to justify a value of $\lambda$
different from its natural value $\hat{M}$.

\section{Quantised fields} 
\label{2}

As we pointed out in our previous paper, quantum effects in scenarios with more
than one extra dimension can be qualitatively different from models without
internal spaces and can, in principle, provide new ways of addressing the
hierarchy. It then seems reasonable to ask the same question in relation to the
class of models described previously.

Therefore we devote this section to the computation of the one-loop effective
potential arising from a massive bulk scalar field $\Phi(x,X,y)$ coupled
non-minimally to the higher dimensional curvature.  We also point out that, as
noted in \cite{garpom,alex}, it is possible to relate the effective potential
from a bulk scalar with the one arising from a gauge field, the computation
being virtually the same. It is possible to do so by appropriately fixing the
non-minimal coupling and the bulk mass of the scalar field to (we take $D_1=4$ and work in the physical gauge)
\bea
\xi &=& 1/8~, \label{3/8}\\
m^2 &=& -k^2/2~.
\label{nuone}
\eea
The field equation for $\Phi(x,X,y)$ is given by the Klein-Gordon equation
\begin{equation}
    \left[-\Box_{D} + m^2 + \xi \hat{{\cal R}}\right] 
    \Phi =0~,
\label{eq3}
\end{equation}
where $\hat{{\cal R}}$ is the higher dimensional curvature and $\Box_{D}$ the
D'Alem\-ber\-tian, both computed from the metric (\ref{metric}), (\ref{type2}).

Standard Kaluza-Klein theory tells us that such a higher dimensional field can
be expressed in terms of a complete set of modes, which describe a tower of
fields with masses quantised according to some eigenvalue problem. Such a
decomposition is, of course, arbitrary, however a convenient choice is
\beq
\Phi(x,X,y) = \sum_{\k,n} \Psi_\k(X)
\varphi_{\k,n}(x)Z_{\k,n}(y)~,
\label{eq6}
\eeq
where the modes $\Psi_\k(X)$ are chosen to be a complete set of solutions of
the Klein-Gordon equation on the manifold $\Sigma$:
\bea
   P_\Sigma \Psi_l(X) &=& 
   \Big[- \Delta_{\Sigma} + \xi 
   {\mathcal R}_{\Sigma}\Big] \Psi_\k(X) \nonumber \\ 
   &=&{1\over R^2}\lambda_\k^2
   \Psi_\k(X)~,
   \label{psigma}
\eea
with eigenvalues $\lambda_\k^2$ (independent of $R$) and degeneracy $g_\k$.  If
we now require $\varphi_{\k,n}(x)$ to satisfy the Klein-Gordon equation in
Minkowski spacetime, $\mathcal M$, with masses $m^2_{\k,n}$,
\beq
  \left[- \Box + m^2_{\k,n}\right]
  \Phi_{\k,n}(x)=0~,
\label{eq8}
\eeq
equation (\ref{eq3}) leaves us with a radial equation for the modes $Z_{\k,n}(y)$ 
\beq
    {\cal D}_y^{(l)} Z_{\k,n} =
    m_{\k,n}^2 Z_{\k,n}~,
    \label{eq10}
\eeq
where the differential operator ${\cal D}_y$ is given by
\beq
{\cal D}_y^{(l)}= e^{2\si}\left[ -e^{-D_1\si} \partial_y e^{D_1\si} 
\partial_y  +
        \mu_l^2   - 2D_1 \xi \si'' \right]~,
\eeq
and
\beq
\mu_l^2 = m^2 + {1\over R^2} \lambda_l^2 - D_1(D_1+1)k^2 \xi~.
\label{eq11}
\eeq
The most general solution to (\ref{eq10}) can be written in terms of Bessel functions and
by imposing the appropriate boundary conditions, we find that the eigenvalues $m_{n}$ are 
determined by the  transcendental equation:
\beq
F_{\nul}^\beta\left({m_{n,l}\over k a}\right)=0~.
\label{masses}
\eeq
The function $F_{\nul}^{\beta}(z)$ is given by
\bea
\label{fs}
F_{\nul}^{\beta}(z) =  Y^\beta_{\nul} (az)J^\beta_{\nul} (z) - J^\beta_{\nul}
(az)Y^\beta_{\nul} (z)~, 
\eea
where
\beq
\nul^2 = {\mu_l^2 \over k^2}+{D_1^2 \over 4}~,
\label{eq13}
\eeq
and 
\beq
J^\beta_{\nul} (z) =  J_{\nul} (z)
\eeq
for twisted field configurations ($Z_{n,l}(-y) = -Z_{n,l}(y)$) or
\beq
J^\beta_{\nul} (z) =  j_{\nul} (z) = {1\over 2}D_1(1-4\xi)J_{\nul}(z) +zJ_{\nul}'(z)~,
\eeq
for untwisted ones ($Z_{n,l}(-y) = Z_{n,l}(y)$). Analogous expressions are valid
also for $Y^\beta_{\nul} (z)$. 
In the following we focus on the case of untwisted fields only.

The one loop effective action $\Gamma^{(1)}$ can be expressed as the sum over
the contributions of each mode \cite{toms}:  
\beq 
\Gamma^{(1)}= - \int d^{4-2\epsilon}x \;V(s)~,
\label{eq15}
\eeq
with
\beq
  V(s) = -{\mu^{2\epsilon}\over 2(4 \pi)^{2}} 
  \Gamma(s) 
  {\sum_{n,l}}' g_l 
  m_{n,l}^{-2s}~,
  \label{eq16}
\eeq 
where the prime in the sum assumes that the zero mass mode is excluded and
$s=-2+\epsilon$.  We are using dimensional regularisation and continuing along
Minkowski spacetime ($4 \rightarrow 4-2\epsilon$) and $\mu$ is a renormalisation
scale introduced for dimensional reasons.

In order to evaluate the sum in (\ref{eq16}), we find convenient to separate the
$\lambda_0-$mode from the rest of the tower\footnote{This procedure is not
  essential, however, by performing such spitting, the RS contribution comes
  about explicitly. Moreover, the RS divergence has to cancel when the two
  contributions are summed and this provides a non-trivial check of the
  calculation}: 
\beq
V(s) = V_{RS}(s) + V_{*}(s)~.
\label{vsplit}
\eeq
The first term corresponds to the usual Randall-Sundrum contribution:
\beq
V_{RS}(s)=- {(ka)^{4}\over 2(4\pi)^2}(ka/\mu)^{-2\epsilon}\Gamma(s){\sum_n}'  g_0
x_{n,0}^{-2s}~, 
\label{vrs}
\eeq
with $x_{n,l} = {m_{n,l} \over ka}$.
This term, present only when $g_0=1$, has been evaluated in 
\cite{flachi1,gpt1} and we report the result without further comments:
\bea
V_{RS}&=&
 -g_0  {k^4 \over 32\pi^2} 
(k/\mu)^{-2\epsilon} 
    \Bigl\{-d_4 {1\over \epsilon}  
        \left(
            1+a^{4-2 \epsilon}\right) \nonumber \\&& 
        + c_1 + a^{4} c_2 - 2 a^4 {\cal V}(a) 
    \Bigr\} ,
\label{rsone} 
\eea
where
\beq 
{\cal V}(a)= \int_0^\infty dz z^3 
\ln\left(1-{k_\nu( z)\over k_\nu( a z)} {i_\nu( a  z)\over i_\nu( z)}\right)
\eeq 
and  the coefficients $c_1$ and $c_2$ do not depend on $a$.
The remaining term in (\ref{vsplit}) is given by
\beq
  V_{*}(s) = -{(ka)^{4}\over 2(4\pi)^2}(ka/\mu)^{-2\epsilon}
  \Gamma(s) 
  \sum_{n,l=1}^{\infty} g_l 
  x_{n,l}^{-2s}~,
  \label{eqv*}
\eeq 
and can be handled in the usual manner by transforming it into a contour
integral and by deforming the contour appropriately, according to a general
technique developed in \cite{lese,klaus1} (See \cite{klausbook} for a
comprehensive review). Standard manipulations lead to 
\bea
V_{*}(s) &=& -{(ka)^{4}\over 2(4\pi)^2}
{(ka/\mu)^{-2\epsilon} \over \Gamma(1-s)} \nonumber \\ &&
\sum_{l=1}^\infty g_l \int_0^\infty dz z^{-2s} {d\over dz}
\ln P_{\nul}(z)
\label{inte0}
\eea
where 
\beq
P_{\nul}(z) = F_{\nul}(iz) = i_{\nul}(az)k_{\nul}(z)-i_{\nul}(z)k_{\nul}(az)~,
\eeq
and
\bea
i_{\nul}(z) &=& zI'_{\nul}(z)+{1\over 2}D_1(1-4\xi)I_{\nul}(z)~,\nonumber \\
k_{\nul}(z) &=& zK'_{\nul}(z)+{1\over 2}D_1(1-4\xi)K_{\nul}(z)~.\nonumber
\eea
Now we have to analytically continue the previous expression (\ref{inte0}) to
the left of $\Re (s) < 1/2$. A possible way of achieving this is to employ the
uniform asymptotic expansion (UAE). This is because the order of the Bessel
function depends explicitly on the eigenvalues $\lambda_l$. 
In order to apply the UAE, we rescale the integral (\ref{inte0}), $z \rightarrow \nul z$:
\bea
V_{*}(s) &=& -{(ka)^{4}\over 2(4\pi)^2}{(ka/\mu)^{-2\epsilon} \over \Gamma(1-s)} \nonumber \\&&
\sum_{l=1}^\infty g_l \int_0^\infty d({\nul}z) ({\nul} z)^{-2s} {d\over d({\nul}
  z)}\ln P_{\nul}({\nul} z)~, 
\label{inte}
\eea
and to isolate the divergent part, we express the integrand as its large
${\nul}$ portion plus terms leading to finite contributions.  

By using (\ref{uaei}), (\ref{uaek}), we can recast (\ref{inte}) as the sum of three terms:
\beq
V_{*}(s) = V_{1}+V_{2}+V_{3}~,
\label{split123}
\eeq
with
\bea
V_{j} &=& -{(ka)^{4}\over 2(4\pi)^2}{(ka/\mu)^{-2\epsilon}\over \Gamma(1-s)} \sum_{l=1}^\infty
g_l \nul^{2s}\nonumber \\ &&
\int_0^\infty dz z^{-2s} {d\over dz} \ln H_j(z)~,
\eea
and 
\bea
H_1(z) &=& 
(1+a^2z^2)^{1/4} e^{-{\nul}\eta(az)}(1+z^2)^{1/4}
    e^{{\nul}\eta(z)}
~, \nonumber 
\eea
\bea
H_2(z) &=& \Sigma_{\nul}^{(I)}(z) \Sigma_{\nul}^{(K)}(az)~, \nonumber
\eea
\bea
H_3(z) &=& 1- e^{2{\nul}(\eta(az)-\eta(z))}
{\Sigma_{\nul}^{(I)}(az) \Sigma_{\nul}^{(K)}(z) \over \Sigma_{\nul}^{(I)}(z)
  \Sigma_{\nul}^{(K)}(az)}~, \nonumber 
\eea
where $\eta(z)$ is defined in the appendix.
The first term is straightforward to evaluate and gives
\bea
V_{1} &=& -{(ka)^{4}\over 8(4\pi)^2}(ka/\mu)^{-2\epsilon} 
\Bigl\{
\Gamma(s)\hat{\zeta}(s)(1+a^{2s}) - \nonumber \\
&&- {1\over 2\sqrt{\pi}} \Gamma(s-1/2) \hat{\zeta}(s-1/2)(1-a^{2s})
\Bigr\}~.
\label{v1}
\eea
The second one is slightly more involved to evaluate. The uniform asymptotic
expansion (\ref{unif1}), (\ref{unif2})  allows us to write 
\bea
V_{2} &=& {(ka)^{4}\over 2(4\pi)^2}(ka/\mu)^{-2\epsilon} \Bigl\{ 
\sum_{n=1}^\infty \sum_{k=0}^n (1+(-1)^na^{2s})  
\nonumber \\&&
\sigma_{n,k}{\Gamma(s+n/2+k)\over  \Gamma(k+n/2)} \hat{\zeta}(s+n/2)
\Bigr\}~. 
\label{v2}
\eea
In order to deal with the sum over the eigenvalues $\nul$, we have defined the
following base $\zeta-$function: 
\beq
\hat{\zeta} (s) = \sum_{l=1}^\infty g_l \nul^{-2s}~.
\label{zed}
\eeq
The last term in (\ref{split123}) is the usual non local contribution and, since
it is finite by construction, we can safely put $\epsilon =0$: 
\beq
V_{3} = {(ka)^{4}\over (4\pi)^2} 
\sum_{l=1}^\infty g_l {\cal V}_l(a) 
\label{v3}
\eeq

where
\beq
{\cal V}_l(a) =  
\int_0^\infty dz z^{3}
\ln \left\{1- {i_{\nul}(az) k_{\nul}(z)
\over i_{\nul}(z) k_{\nul}(az)}
\right\}~.
\eeq
In order to make the $R-$dependence in (\ref{v1}) and (\ref{v2}) explicit, it is
convenient to rescale the above defined $\zeta-$function by expanding the
binomial. A simple calculation gives: 
\beq
\hat{\zeta} (s) = {\left(kR\right)^{2s} \over \Gamma(s)} \sum_{q=0}^\infty
{(-1)^q\over q!}(kR\nu)^{2q} 
\Gamma(s+q) \zeta(s+q)
\label{zetahat}
\eeq
where
\beq
\zeta(s) = \sum_{l=1}^\infty g_l \lambda_l^{-2s}
\label{zeta}
\eeq
does not depend on $R$ and 
$$
\nu^2 = {m^2\over k^2} - D_1(1+D_1)\xi + {D_1^2 \over 4}~.
$$
The use of (\ref{zetahat}) allows us to express the result in terms of the
generalised $\zeta-$function (\ref{zeta}) and the additional (Mittag-Leffler)
representation for the $\zeta-$function can then be used to deal with the pole
structure of (\ref{zeta}) and express the residues at the poles in terms of
geometrical quantities \cite{blau}.  The Mittag-Leffler representation for the
$\zeta-$function associated with the operator $P_\Sigma$ (see, for example,
\cite{flachi4}) is
\beq
\zeta(s) = {1\over \Gamma(s)}\left\{ \sum_{p=0}^\infty {\tilde{C}_p \over s-D_2/2+p} + f(s)\right\}~,
\label{ML}
\eeq
where $\tilde{C}_p=C_p - g_0 \delta_{p,D_2/2}$ and the $C_p$ are the heat-kernel
coefficients of the operator $P_\Sigma$, $p$ runs over the positive half
integers and $f(s)$ is an entire function. As in the case of \cite{flachi4},
since the internal space $\Sigma$ is boundaryless the heat-kernel coefficients
of semi-integer order are zero.  Relation (\ref{ML}) can now be used to regulate
the effective potential and some calculations lead to
\bea
V(s)&=&
{(ka)^{4}\over 2(4\pi)^2}(ka/\mu)^{-2\epsilon} \left(kR\right)^{2s}  \nonumber \\&& 
\sum_{n=-1}^\infty \sum_{q=0}^\infty (1+(-1)^na^{2s}) 
\left({1\over \epsilon} {\bf a}_{n,q} + {\bf b}_{n,q}\right)(kR)^{2q+n}  \nonumber \\&&
-{g_0 k^4 \over  2(4\pi)^2} (k/\mu)^{-2\epsilon} \left\{ -{d_4\over
      \epsilon}\left(1+a^{4-2 \epsilon}\right) + c_1 + a^{4} c_2 \right\}
\nonumber \\&& 
+ {(ka)^{4}\over (4\pi)^2} \left\{ g_0 {\cal V}(a)
+ \sum_{l=1}^\infty g_l {\cal V}_l(a) \right\}
\label{unrenpot}
\eea
where the coefficients of the previous expression can be written as
\beq
{\bf a}_{n,q} = {(-1)^q\over q!}\nu^{2q} \tilde{C}_{2+D_2/2-n/2-q} {\bf A}_n
\eeq
where
\bea
{\bf A}_{-1} &=& {1\over 8\sqrt{\pi}}~,\nn
{\bf A}_0 &=& -{1\over 4}~,\nn
{\bf A}_n &=& \sum_{k=0}^n {S}_{n,k}~,~~ \mbox{for} ~n>1\nn
{S}_{n,k} &=& {\Gamma(k+n/2+s)\over \Gamma(k+n/2)\Gamma(n/2+s)} \sigma_{n,k}
\eea
The coefficients ${\bf b}_{n,q}$ are related to the ${\bf a}_{n,q}$ via the following correspondence 
$$
{\bf b}_{n,q}= {\bf a}_{n,q}(\tilde{C}_p \rightarrow \Omega_{-p})~,
$$
where the $\Omega_p$ represent the finite part in the power series of 
$\Gamma(s)\zeta(s)$ around $s=p$.

A check on the previous result is provided by the cancellation of the (lower
dimensional) RS divergence, given by
\bed
g_0d_4 {k^4(1+a^4)\over 32\pi^2\epsilon} =
g_0k^4{(1+a^4) \over 32\pi^2\epsilon} (\Delta_0 + \Delta_2 \nu^2 + \Delta_4 \nu^4)~,
\eed
where
\bea
\Delta_0 &=& -{27\over 128}+{3\over 8}\Delta-{1\over 2}\Delta^2+{1\over 2}\Delta^3-{1\over 4}\Delta^4\nn 
\Delta_2 &=& {13\over 16}-\Delta+{1\over 2}\Delta^2\nn 
\Delta_4 &=& -{1\over 8}\nn 
\Delta &=& {1\over 2}D_1(1-4\xi) \label{D}~. 
\eea
A simple inspection of (\ref{unrenpot}) shows that the relevant terms for such a
cancellation are the ones corresponding to  
the couples $(n,q)=(0,2)~,(2,1)~,(4,0)$. Such terms can be easily extracted from
(\ref{unrenpot}) and the use of the coefficients $\sigma_{n,k}$ (the relevant
ones are reported in the appendix, (\ref{sigs})) shows that the RS divergence is
indeed cancelled.

The result for the vacuum energy (\ref{unrenpot}) is divergent and needs to be renormalised. 
The counterterm action can be constructed analogously to the case of two equal warpings \cite{flachi4}:
\bea
S_{n,q} &=& {1\over 32\pi^2 \epsilon} 
\sum_\pm \int d^Dx 
\sqrt{\hat g_\pm} \,\hat\kappa_\pm^{(n,q)} \;\hat{\cal R}_\pm^{(4+D_2-n-2q)/2} 
= \nonumber \\ 
&=& {1\over 32\pi^2 \epsilon}\int d^{4-2\epsilon}x \;{(kR)^{2q+n}\over R^4}
\left\{ a^{4-2\epsilon} + (-1)^n \right\} \kappa^{(n,q)} \nonumber
\eea
where we have defined (the factor proportional to $v_\Sigma$ has been reabsorbed
in the coefficients $\kappa^{(n,q)}$) 
\beq
\hat\kappa^{(n,q)}_-=(-1)^n \hat\kappa^{(n,q)}_+ =k^{2q+n}\kappa^{(n,q)}~,
\eeq
and it is easy to see that all the divergences can be reabsorbed in counterterms of the previous type.
Once we subtract the counter-terms, we arrive at the following expression for
the renormalised effective potential 
\bea
V(a)&=& 
{(1/R)^{4}\over 2(4\pi)^2}
\sum_{n=-1}^\infty \sum_{q=0}^\infty  
{{\bf a}_{n,q} \ln(\mu R)^2 + {\bf b}_{n,q}
\over (a^4+(-1)^n )^{-1}} \; (kR)^{2q+n} \nonumber \\&& 
-{g_0 k^4 \over  2(4\pi)^2} \left[ c_1 + a^{4} c_2 + (1+a^4)d_4\ln(k/\mu)^2 \right] \nonumber \\&&
+ {(ka)^{4}\over (4\pi)^2} \left[ g_0 {\cal V}(a)
+ \sum_{l=1}^\infty g_l {\cal V}_l(a) \right]~.
\eea

\section{Radion stabilisation and quantum self-consistency of the solution}
\label{4}

In the previous section we have computed and renormalised the Casimir energy
arising from a massive bulk scalar field non-minimally coupled to the curvature
and from a massless bulk gauge field.  So we are now in the position to see
whether or not quantum effects provide a reasonable stabilisation mechanism for
the class of models of the type $AdS \times \Sigma$.  To this aim, let us
consider the full action $S$, where we include the contribution $\Gamma^{(1)}$
arising from a quantised field:
\beq
S= S_{BG} + \Gamma^{(1)} ,
\label{full}
\eeq
where we generically write the quantum contribution as
\beq
\Gamma^{(1)} = -\int d^4x \sqrt{\tilde{g}} V(a) .
\eeq
$S_{BG}$ is the classical background action obtained by using the ansatz for the
metric (\ref{metric}) (with  
$\eta_{\mu\nu} \rightarrow \tilde{g}_{\mu\nu}(x)$) in (\ref{bulk}) and by
integrating out the extra $D_2+1$ dimensions.  
Now, varying the full action $S$ with respect to $\tilde{g}_{\mu\nu}(x)$
\beq
{\delta S \over \delta \tilde{g}_{\mu\nu}}= 0~,
\label{scsol}
\eeq
and requiring that the minimum is at $\tilde{g}_{\mu\nu}(x)= \eta_{\mu\nu}$ will
tell us whether  or not the classical solution is spoiled by quantum effects. On
the other side, varying $S$ with respect to the radion $a$ 
\beq
{\delta S \over \delta a}= 0~,
\label{a}
\eeq
at $\tilde{g}_{\mu\nu}(x)= \eta_{\mu\nu}$, will tell us whether we can obtain an
exponentially large hierarchy, $a=e^{-\pi kr}$ (with $kr \sim 12$), in which
case such solution also solves the hierarchy problem. We want to stress that one
can have solutions that satisfy (\ref{scsol}) but not (\ref{a}) and therefore
are self-consistent but do not solve the hierarchy problem.  
A simple calculation shows that by combining the previous requirements
(\ref{scsol}), (\ref{a}), the following constraint for the effective potential
is obtained: 
\beq
(1-a^4)V'(a) + 4a^3V(a)=0~,
\label{stabi}
\eeq
where the prime denotes derivative with respect to $a$. Equation (\ref{stabi})
is exactly the same as the one obtained for the RS model \cite{flachi1}. 

We have now to specify the matter content of our model and in turn the function
$V(a)$. We consider two possibilities: 
a minimal model whose action is given by (\ref{bulk}) and a non-minimal model
where (\ref{bulk}) is our classical background theory upon which we quantise a
bulk scalar field. In the first case, we assume that the gauge field splits into
a classical plus a quantum part:  
\beq
A_\mu = A_{\mu}^{C} + A_{\mu}^{Q}~,
\eeq
and thus the quantum contribution comes from the quantum counterpart of the
gauge field.  We shall consider the $AdS$ components only, which have a zero vev
and do not couple to the Yang-Mills flux configuration.  In the non-minimal
case, it is the scalar field that provides us with the quantum effective
potential.

We recast the result for the effective potential as follows:
\beq
V(a) = {k^4 \over 32 \pi^2}\Bigl\{
\Gamma_1 + a^4\Gamma_2
 + \Gamma_{NL}(a) \Bigr\}
\label{potpot}
\eeq
where $\Gamma_1$ and $\Gamma_2$ do not depend on $a$. The non-local contribution,
\beq
\Gamma_{NL}(a) = a^4 {\cal V}(a) + a^4 \sum_{l=1}^\infty g_l
{\cal V}_{l}(a)~,
\label{gammanl}
\eeq
is slightly more involved to inspect, however, in our case, it is sufficient to
see that the contribution coming from the massive Kaluza-Klein modes (involving
the sum over $l$) is highly suppressed with respect to the (RS) zero-mode term,
proportional to ${\cal V}(a)$. This can be shown by noticing that the dominant
contribution to the integral in ${\cal V}_{l}(a)$ comes from the region $z
\lesssim 1$. Expanding the integrand in such region allows one to see that
${\cal V}_{l}(a)$ goes like $a^{2\nul}$ and a simple inspection of the sum tells
us that the non-local contribution coming from the massive KK states is
proportional to powers of $a^{1/(kR)}$.
The non local contribution can then be approximated as
\beq
\Gamma_{NL}(a) \simeq a^4 {\cal V}(a)~.
\eeq
Fixing the field content of the theory (or the bulk parameters) will uniquely
determine the function ${\cal V}(a)$. 
(Such term has been evaluated for any of $\nu$ in \cite{garpom}). By expanding
the integrand for small $a$, one finds that in the minimal case (only with a
quantised gauge field)  
\beq
{\cal V}(a) = {\gamma \over \ln a}~,
\label{nonloclog}
\eeq
with $\gamma$ being $a-$independent. In the non-minimal case one has to
distinguish  three possibilities: when the order of the Bessel functions
is $\nu=0$, this corresponds to taking  
\beq
\xi = {4m^2+D_1^2k^2\over 4D_1(D_1+1)k^2} ,
\label{nuzero}
\eeq
when $\nu=1$ and this corresponds to fixing the values of $\xi$ and $m$
according to (\ref{3/8}) and (\ref{nuone}), and finally, when $\nu$ is different
from the two previous values\footnote{
  Fixing the bulk matter content can also be understood as a sort of tuning,
  which can be removed only by a more fundamental theory that leads to the
  specified field content. Moreover, special values of the mass of bulk scalars
  are unstable under quantum corrections unless supersymmetry is present}.
In the first case, we find
\beq
{\cal V}(a) = {\beta \over \alpha + \ln a}~,
\eeq
where $\alpha$ and $\beta$ do not depend on $a$.  The second case, obviously,
gives back relation (\ref{nonloclog}), whereas in the third case ${\cal V}(a)$
is proportional to $a^N$ with $N \geq 4$.

The previous relations along with the self-consistency condition (\ref{stabi})
allow us to see in which cases we obtain a solution to the hierarchy problem
with the bonus for the solution to be self-consistent.

By using the expression for the effective potential (\ref{potpot}) and
(\ref{nonloclog}), we find that the solution to (\ref{stabi}) in the minimal
case, in the limit of $a\ll1$, is
\beq
a \sim e^{-{\gamma / (\Gamma_1 + \Gamma_2})}
\eeq
which shows that there is no need of any fine tuning in order to get an exponentially small $a$.

In the non-minimal case, one can easily check that fixing the parameters $\xi$
and $m$ according to (\ref{nuone}) or (\ref{nuzero}) provides also a large
hierarchy, whereas in the other cases no solution to (\ref{stabi}) is found for
small values of $a$.

\section{Conclusions}
\label{5}

In this article we have investigated the role of quantum effects
arising from bulk fields in higher dimensional brane
models. Specifically, we have considered a class of warped brane
models whose topology is $AdS_5 \times \Sigma$,
where $\Sigma$ is a $D_2$ dimensional one-parameter compact
manifold and two branes of codimension one are placed
at the orbifold fixed points. 

We have seen that such a set-up can be obtained from Einstein-Yang-Mills theory.
Contrarily to the case studied in \cite{flachi4}, where both the radion $a$ and
the radius of $\Sigma$ were undetermined classically, here the radius of the
internal space $\Sigma$ is stabilised at a size comparable with the higher
dimensional cut-off once the Yang-Mills flux is tuned according to (\ref{effe}).
This guarantees that, when matter is placed on the wall, the extra dimensions in
the $\Sigma -$direction remain invisible, as it must be. On the other hand, the
fact that the size of the internal manifold is of order $1/\hat{M}$, does not
suggest any new way of addressing the hierarchy, which is resolved only through a
redshift effect coming from the $AdS$ direction.

We considered two possible scenarios: a first one, which we labeled as
`minimal', where the action is just the Einstein-Yang-Mills one and only the
Yang-Mills field is quantised, and a second one, called `non-minimal', in which
we quantised a scalar field on a classical Einstein-Yang-Mills background.

We evaluated the renormalised one-loop effective action at lowest order, namely
the Casimir energy in the case of a massive non-minimally coupled scalar field.
The resulting scalar effective potential can be related to the one arising from
quantised gauge fields by appropriately fixing the parameters $\xi$ and $m$.

The computation is similar to the one carried out in \cite{flachi4}, with some
technical differences due to the explicit presence of the eigenvalues of the
scalar operator on the manifold $\Sigma$ in the order of the Bessel functions.
This can be effectively dealt with by using the uniform asymptotic expansion of
the modes, which turned out slightly more involved than the corresponding
computation in the case of \cite{flachi4}. On the other hand, the Mittag-Leffler
expansion for the generalized $\zeta-$function allowed us to express the Casimir
energy in terms of heat-kernel coefficients of the internal space $\Sigma$ as in
the case previously considered in \cite{flachi4}. The same non trivial check of
the cancellation of the RS divergence works. Also the renormalisation is carried
out analogously to \cite{flachi4} by subtracting suitable counter-terms
proportional to a number of boundary or bulk local operators.

Finally we investigated the self consistency of the model by requiring that the
quantum corrected action is minimised by the background solution. As for the
stabilization our analysis indicates that the Casimir force can stabilise the radion without
fine tuning thanks to any KK mode whose index $\nul$ is $0$ or
$1$. The latter is reproduced in the `minimal' case by the zero mode of the gauge field, 
in analogy to what was found in \cite{garpom} in the Randall
Sundrum context.  Obviously, the (scalar) laplacian $\Delta_\Sigma$ defined on
$\Sigma$ must have a zero eigenvalue for this to happen.

For scalar fields, a large contribution to the effective potential is produced only at the price of 
choosing appropriately the mass and the non-minimal coupling.
In such case a large hierarchy is generated, but the masses of the modes are unstable under quantum effects and supersymmetry has to be invoked.

It would be interesting to apply the previous ideas to models where both 
the $x-$
and $X-$coordinates cover curved internal spaces, as it could, for example,
happen in a cosmological scenario. In such case the self-consistency condition
is more involved than the one presented here and might allow more interesting
conclusions.

\acknowledgements 

The constant advise of J. Garriga throughout the completion of this work is
mostly appreciated.  A.F. acknowledges fruitful discussions with A.  Pomarol, L.
Da Rold and W. Naylor. The work of A.F. is supported by the European Community
via the award of a Marie Curie Fellowship. The support of FEDER under project
FPA 2002-00748 is also kindly acknowledged. O.P. is supported by MICYT under grant
FPA 2002-00648 and by DURSI under grant 2001SGR00188.

\appendix
\section{Uniform Asymptotic Expansion and computation of the coefficients $\sigma_{n,k}$} 
\label{sigmas}

In the present appendix, we report the relevant formulas concerning the uniform asymptotic expansion (UAE) for the altered Bessel functions, $i_{\nul}(z)$ and 
$k_{\nul}(z)$, used in the computation of $V_2$. By using the results reported in 
\cite{klausbook,abramowitz}, we find
\bea
i_{\nul}(\nul z) &=& {\nul e^{\nu_l \eta}\over \sqrt{2\pi \nul}} (1+z^2)^{1/4}
\Sigma_{\nul}^{(I)}(z)~,\label{uaei}\\ 
k_{\nul}(\nul z) &=& - \sqrt{{\pi \nul \over 2}}  e^{\nul \eta} (1+z^2)^{1/4} \Sigma_{\nul}^{(K)}(z)~,
\label{uaek}
\eea
with
\beq
\Sigma_{\nul}^{(I)}(z)={1\over 2\nul \sqrt{1+z^2}}D_1(1-4\xi) \Sigma_1 + \Sigma_2~,
\eeq
and
\beq
\Sigma_{\nul}^{(K)}(z)={1\over 2\nul \sqrt{1+z^2}}D_1(1-4\xi) \Sigma_3 - \Sigma_4~,
\eeq
where
\bea
t &=& {1\over \sqrt{1+z^2}}~,\nonumber \\ 
\eta(z)&=& \sqrt{1+z^2} + \ln \left({z\over 1 + \sqrt{1+z^2}}\right)~. \nonumber 
\eea
The functions $\Sigma_I$ are given by
\bea
\Sigma_1 &=& \sum_{k=0}^\infty {u_k\over \nul^k} ~,\nonumber \\ 
\Sigma_2 &=& \sum_{k=0}^\infty {v_k\over \nul^k} ~,\nonumber \\ 
\Sigma_3 &=& \sum_{k=0}^\infty (-1)^k {u_k\over \nul^k} ~,\nonumber \\ 
\Sigma_4 &=& \sum_{k=0}^\infty (-1)^k{v_k\over \nul^k} ~.\nonumber
\eea
with the coefficients of the previous expansions expressed by the following recursion relations:
\bea
u_{k+1}(t) &=& {1\over 2} t^2 (1-t^2) u'_{k}(t)+{1\over 8}\int_0^t (1-5x^2)u_k(x)dx \nonumber \\
v_{k+1}(t) &=& u_{k+1}(t)- {1\over 2}t(1-t^2)u_{k}(t)-t^2(1-t^2)u'_k(t) \nonumber
\eea
with $u_0(t)=1$. 
It is possible to expand the previous functions in powers of $\nul$:
\beq
\Sigma_{\nul}^{(I)}(z) = 1 + \sum_{j=1}^\infty {p_j(t)\over \nul^j}~,
\eeq
\beq
\Sigma_{\nul}^{(K)}(z) = 1+ \sum_{j=1}^\infty (-1)^j {p_j(t)\over \nul^j}~,
\eeq
where
\beq
p_j(t)= {D_1 (1-4\xi)\over 2} tu_{j-1} +v_j~.
\eeq
It is now easy to see that, in order to obtain the coefficients $\sigma_{n,k}$,
we only need to expand the logarithm of $\Sigma_{\nul}^{(I,K)}(z)$: 
\bea
\ln \left( 1 + \sum_{j=1}^\infty {p_j(t)\over \nul^j}
\right) &=& 
\sum_{n=1}^\infty \sum_{k=0}^n \sigma_{n,k} {t^{n+2k}\over \nul^n}~.
\label{unif1}\\ 
\ln \left(1 + \sum_{j=1}^\infty (-1)^j {p_j(t)\over \nu^j}
\right) &=& 
\sum_{n=1}^\infty \sum_{k=0}^n (-1)^n\sigma_{n,k} {t^{n+2k}\over \nul^n}~. 
\label{unif2}
\eea
The coefficients $\sigma_{n,k}$ can be obtained by using any symbolic
manipulation program. We report here only the ones needed to cancel the RS
divergence: 
\bea
\sigma_{4,0} &=& -{27 \over 128}+{3\over 8}\Delta - {1\over 2}\Delta^2 + {1\over
  2}\Delta^3 - {1\over 4}\Delta^4~, \nn 
\sigma_{2,1} &=& {5\over 8}-{1\over 2}\Delta~, \nn
\sigma_{2,0} &=& -{3 \over 16}+{1\over 2}\Delta - {1\over 2}\Delta^2~,
\label{sigs}
\eea
with $\Delta$ given by (\ref{D}).

\vfill \eject

\end{document}